\documentclass[twocolumn,preprintnumbers,amsmath,amssymb]{revtex4}
\usepackage{graphicx}
\usepackage{dcolumn}
\usepackage{bm}

\begin{document}

\title{Polaron Mass and Electron-Phonon Correlations in the Holstein Model }

\author{ Marco Zoli }
\affiliation{
Dipartimento di Fisica - Universit\'a di Camerino, I-62032, Italy. - marco.zoli@unicam.it}

\date{\today}

\begin{abstract}
The Holstein Molecular Crystal Model is investigated by a strong coupling perturbative method which, unlike the standard Lang-Firsov approach, accounts for retardation effects due to the spreading of the polaron size. The effective mass is calculated to the second perturbative order in any lattice dimensionality for a broad range of (anti)adiabatic regimes and electron-phonon couplings. The crossover from a large to a small polaron state is found in all dimensionalities for adiabatic and intermediate adiabatic regimes. The phonon dispersion largely smooths such crossover which is signalled by polaron mass enhancement and on site localization of the correlation function. The notion of self-trapping together with the conditions for the existence of light polarons, mainly in two- and three-dimensions, are discussed. By the imaginary time path integral formalism I show how non local electron-phonon correlations, due to dispersive phonons, renormalize downwards the {\it e-ph} coupling justifying the possibility for light and essentially small 2D Holstein polarons.
\end{abstract}

\maketitle

\section*{1. Introduction}

The interest for phonons and lattice distortions in High Temperature Superconductors (HTSc) is today more than alive \cite{i37,i33}.
While the microscopic origin of the pairing mechanism in cuprate HTSc has not yet been unravelled, evidence has been provided \cite{i2,i4,i38,i5,i39,i6} that electron-lattice interactions and local lattice fluctuations are correlated with the onset of the superconducting transition \cite{i3,i34,i121}.  Recent investigations \cite{i9} would suggest a similar role for the lattice also in the newly discovered layered pnictide-oxyde quaternary superconducting compounds \cite{i7,i8,i119}.  In fact, the discovery of HTSc in copper oxides \cite{i10} was motivated by the knowledge that the Jahn-Teller effect \cite{i11,i12,i36} is strong in $Cu^{2+}$. In nonlinear molecules a lattice distortion lifts the degeneracy of the electronic states and lowers the overall ground state energy: the energy gain is the Jahn-Teller stabilization energy which can be measured optically. As vibrational and electronic energies are of the same order, the nuclear motion cannot be separated from the electrons and the combined electron-lattice object becomes a mobile Jahn-Teller polaron  \cite{i13} which can be described in terms of the nonlinear Holstein Hamiltonian \cite{i14}.
Over the last twenty years, the focus on the HTSc has largely contributed to trigger the study of the polaron properties \cite{i15,i72,i35,i0}.  Earlier and more generally, polarons had become a significant branch in condensed matter physics \cite{i40,i115,i41,i71} as Landau introduced the concept of an electron which can be trapped by digging its own hole in an ionic crystal \cite{i16}. A strong coupling of the electron to its own lattice deformation implies violation of the Migdal theorem \cite{i99} and polaron collapse of the electron band \cite{i100} with the appearance
of time retarded interactions in the system. A great advance in the field came after Feynman used the path integrals
\cite{i73} to calculate energy and effective mass of the Fr\"{o}hlich polaron \cite{i42} by a variational method \cite{i43}. In the statistical path integral, the quantum mechanical electron motion is represented by a real space path ${\bf x}(\tau)$ being function of an imaginary time $\tau$  whose range has an upper bound given by the inverse temperature. As a key feature of the Feynman treatment, the phonon degrees of freedom can be integrated out exactly (being the action quadratic and linear in the oscillator coordinates) but leave a substantial trace as a time retarded potential naturally emerges in the exact integral action. {\it The electron, at any imaginary time, interacts with its position at a past time}. This self interaction mirrors the fact that the lattice needs some time to readjust to the deformation induced by the electron motion and follow it. For decades, the Fr\"{o}hlich polaron
problem has been extensively treated by path integral techniques \cite{i44,i74} which, remarkably, can be applied for any value of the coupling constant \cite{i118}. A review work on the Fr\"{o}hlich polaron is in Ref. \cite{i120}.

With regard to small polarons,  path integral investigations started with the groundbreaking numerical work  by De Raedt and Lagendijk \cite{i18,i95} who derived fundamental properties for the Holstein polaron.
While a sizable electron-phonon coupling is a requisite for polaron formation \cite{i50,i116,i117}, also the dimensionality and degree of adiabaticity of the physical system could essentially determine the stability and behavior
of the polaron states \cite{i17,i19,i20,i79,i70,i32,i22,i77,i23,i93,i80,i25,i26,i60,i92}. When the characteristic phonon energy  is smaller than the electronic energy  the system is set in the adiabatic regime. In materials such as the HTSc, colossal magnetoresistance oxides \cite{i31,i83}, organic molecular crystals  \cite{i47}, DNA molecules \cite{i110,i111} and finite quantum structures \cite{i108,i109},  intermediate adiabatic polarons  are relevant as carriers are strongly coupled to high energy optical phonons.
For the HTSc systems, a path integral description had been proposed for polaron scattering by anharmonic
potentials, due to lattice structure instabilities, as a possible
mechanism for the non metallic behavior of the c-axis resistivity \cite{i45,i46}.

It has been questioned whether small (bi)polarons could indeed account for high $T_c$  due to their large effective mass \cite{i28} but, later on, it was recognized that dispersive phonons renormalize the effective coupling in the Holstein model yielding much lighter masses \cite{i24}. Also long range Fr\"{o}hlich e-ph interactions provide a pairing mechanism and bipolaron mass consistent with high $T_c$ \cite{i29,i48,i61,i84} while accounting, among others, for angle-resolved photoemission spectroscopy data in cuprates HTSc \cite{i82,i49}.

In view of the relevance of the polaron mass issue,  I review in this article some work \cite{i27} on the Holstein polaron {\it with dispersive phonons} which examines the notion of self-trapping and estimates some polaron properties in the parameter space versus lattice dimensionality. The latter is introduced in the formalism by modelling the phonon spectrum through a force constant approach which weighs the first neighbors intermolecular shell. This is done in the spirit of the work by Holstein \cite{i86} on the small polaron motion for a one dimensional narrow band system. Calculating  the polaron hopping between localized states due to multiphonon scattering processes in perturbation theory, Holstein first pointed out that the phonon dispersion was in fact {\it a vital ingredient of the theory} whereas a dispersionless spectrum would yield a meaningless divergent hopping probability.
Time dependent perturbative theory, the hopping integral being the perturbation, applies as long as the {\it interaction time} is shorter than the polaron lifetime on a state.
Computing  the time integral for the hopping probability, I had earlier shown \cite{i32} that such interaction time is shorter in higher dimensionality thus making the time dependent perturbative method more appropriate in the latter conditions. For a given dimensionality, the interaction time is also reduced by increasing the strength of the intermolecular forces hence, the phonon dispersion.
Consistently also the crossover temperature between band-like motion (at low $T$) and hopping motion (at high $T$) \cite{i91}  shifts upwards, along the $T$ axis, in higher dimensional systems with larger coordination numbers. Hopping polaron motion is the relevant transport mechanism in several systems including DNA chains \cite{i112,i113}.

It is worth emphasizing that our previous investigations on the Holstein polaron and the present paper assume dimensional effects as driven by the lattice while the electron transfer integral is a scalar quantity. In a different picture, first proposed by Emin \cite{i87} for quasi 1D solids in the adiabatic limit and recently generalized by variational exact diagonalization techniques \cite{i88}, the electron transfer integral is instead a vector whose components are switched on in order to treat anisotropic polaron properties. The system dimensionality is then increased by tuning the electronic subsystem parameters, the 3D case corresponding to an isotropic hopping integral, whereas the phonon spectrum is dispersionless. While the method to introduce the system dimensionality in Ref. \cite{i88} differs from ours, some trends regarding the polaron mass in 1D and 3D are apparently at variance with ours as it will be pointed out in the following discussion.

Being central to polaron studies, the effective mass behavior ultimately reflects the abovementioned property of the polaron problem: when the electron moves through the lattice, it induces and drags a lattice deformation which however does not follow instantaneously, the retardation causing a spread in the size of the quasiparticle. The electron-phonon correlation function provides a measure of the deformation around the instantaneous position of the electron and may be used to quantify the polaron size. Several refined theoretical tools, including quantum Monte Carlo calculations \cite{i18}, density matrix renormalization group \cite{i21}, variational methods \cite{i75,i94} and exact diagonalization techniques \cite{i90,i96} have been applied to the Holstein Hamiltonian to compute such polaron characteristics covering various regimes of electron-phonon coupling and adiabatic ratio. While a qualitative and often quantitative \cite{i21,i75} agreement among several numerical methods has emerged especially with regard to ground state polaron properties, we feel that analytical investigations may provide a useful insight also in the specific regime of intermediate adiabaticity and e-ph coupling for which the leading orders of  perturbation theory traditionally fail in the weak coupling \cite{i97} or become less accurate in the strong coupling \cite{i53,i98}.

Extremely interesting is that range of e-ph couplings which are sufficiently strong to allow for polaron formation but not too strong to prevent the polaron from moving through the lattice. When the relevant energy scales for electrons and phonons are not well separated, such range of couplings produces a composite particle which is not trapped by its lattice deformation. It is this intermediate regime which forms the focus of the present work.

The retardation effect and its consequences are here treated by means of a variational analytical method based on the Modified Lang Firsov (MLF) transformation \cite{i51} which considerably improves the standard Lang Firsov (LF) \cite{i30} approach on which strong coupling perturbation theory (SCPT) is based. Section 2 provides the generalities of the MLF method applied to the dispersive Holstein Hamiltonian. In Section 3, the  polaron mass is calculated in the (anti)adiabatic regimes covering a broad range of cases in parameter space. The spreading of the adiabatic polaron size over a few lattice sites is shown in Section 4 through computation of the electron-phonon correlations. In Section 5, I give further physical motivations for the existence of light Holstein polarons: using a path-integral method, I show how the electron-phonon coupling is renormalized downwards in momentum space by nonlocal correlations arising from the dispersive nature of the phonon spectrum. Some final remarks are in Section 6.

\section*{2. Modified Lang-Firsov Method }

The Holstein diatomic molecular model was originally cast \cite{i14} in the form of a discrete nonlinear Schr\"{o}dinger equation for electrons whose probability amplitude at a molecular site depends on the interatomic vibration coordinates \cite{i101}. The nonlinearities are tuned by the electron-phonon coupling $g$, whose strength drives the crossover between a {\it large} and a {\it small} polaron for a given value of the adiabatic parameter. The polaron radius is measured with respect to the lattice constant \cite{i52,i89}. When the size of the lattice distortion is of the same order of (or less than) the lattice
constant the polaron has a small radius and the discreteness of the lattice must be taken into account.

In second quantization the dimension dependent Holstein Hamiltonian with dispersive harmonic optical phonons reads:

\begin{eqnarray}
H =\, - t \sum_{<i j>} c_i^{\dag} c_{j}
+  g  \sum_i n_i (b_i^{\dag} + b_i)
+ \sum_{\bf q} \omega_{\bf q} b_{\bf q}^{\dag} b_{\bf q}
\label{eq:1}
\end{eqnarray}

$t$ is the hopping integral and the first sum is over $z$ nearest neighbors. $c_i^{\dag}$ and
$c_i$ are the real space electron creation and annihilation
operators at the $i$-site, $n_i \,( = c_i^{\dag} c_i)$ is the number operator,
$b_i^{\dag}$ and $b_i$ are the phonons creation and annihilation
operators. $b_{\bf q}^{\dag}$ is the Fourier transform of
$b_i^{\dag}$ and $\omega_{\bf q}$ is the frequency of the phonon with vector momentum ${\bf q}$.
Unlike the Su-Schrieffer-Heeger Hamiltonian \cite{i102,i103} a paradigmatic model in polymer physics, in Eq.~(\ref{eq:1}) the electron hopping does not depend on the relative displacement between adjacent molecular sites hence, the phonon created by $b_i^{\dag}$ is locally coupled to the electronic density.

The LF transformation uses a phonon basis of fixed
displacements (at the electron residing site) which diagonalizes the Hamiltonian in Eq.~(\ref{eq:1}) in absence of
hopping. The hopping term is then treated as a perturbation \cite{i53,i24}.
However the standard LF approach does not
account for the retardation between the electron and the lattice
deformations which induces a
spread in the size of the polaron. Precisely this effect becomes important for the
intermediate {\it e-ph} coupling values which may be appropriate for some HTSc.

The idea underlying the MLF transformation \cite{i54} is that to consider the displacements of the oscillators {\it at different sites} around an electron in order to describe the retardation effect.
For the present case of dispersive phonon the MLF transformation, applied to the
Hamiltonian in Eq.~(\ref{eq:1}), reads:

\begin{eqnarray}
& &\tilde{H} =\, e^S H e^{-S} \, \nonumber
\\
& &S =\,  \sum_{\bf q} \lambda_{\bf q} n_{\bf q}  (b_{-{\bf
q}}^{\dag} - b_{\bf q}), \nonumber
\\
& &n_{\bf q} =  {1 \over {\sqrt N}}\sum_i n_i e^{-i{\bf q} \cdot {\bf R_i}}
= {1 \over {\sqrt N}}\sum_{\bf k} c_{\bf k+q}^{\dag} c_{\bf k} \nonumber
\\
\label{eq:2}
\end{eqnarray}

where  ${\bf R_i}$ are the lattice vectors and $\lambda_{\bf q}$ are the variational parameters which
represent the shifts of the equilibrium positions of the
oscillators (quantized ion vibrations) with momentum ${\bf q}$.
The conventional Lang-Firsov transformation is recovered by setting: $\lambda_{\bf q}=\, g/\omega_{\bf q}$.

Explicitly, the MLF transformed Holstein Hamiltonian  in Eq.~(\ref{eq:2})
is:

\begin{eqnarray}
& &\tilde{H} =\, - \epsilon_p \sum_i n_i
- t_p \sum_{{i j}} c_i^{\dag} c_{j}\,
\nonumber \\
& &\times \, {\exp}[{1 \over {\sqrt N}}
\sum_{\bf q} \lambda_{\bf q} b_{\bf q}^{\dag}(e^{i{\bf q}
\cdot {\bf R_i}} - e^{i{\bf q} \cdot {\bf R_i}})]
\nonumber \\
& &\times \, {\exp}[-{1 \over {\sqrt N}}
\sum_{\bf q} \lambda_{\bf q} b_{\bf q} (e^{-i{\bf q}
\cdot {\bf R_i}} - e^{-i{\bf q} \cdot {\bf R_i}})]
\nonumber \\
& & + \sum_{\bf q} \omega_{\bf q} b_{\bf q}^{\dag} b_{\bf q}
+ \sum_{\bf q} (g - \lambda_{\bf q} \omega_{\bf q}) n_{\bf q} (b_{-{\bf q}}^{\dag} + b_{\bf q})
\,
\nonumber \\
\label{eq:3}
\end{eqnarray}

where the polaron self-energy $\epsilon_p$ is:

\begin{eqnarray}
\epsilon_p = {1 \over N}~\sum_{\bf q} (2 g - \lambda_{\bf q}
\omega_{\bf q}) \lambda_{\bf q}
\label{eq:4}
\end{eqnarray}

and the polaronic hopping is:

\begin{eqnarray}
& &t_p = t \exp \bigl[ -{1 \over N}
\sum_{\bf q} \lambda_{\bf q}^2 (1- {{\gamma_{\bf q}} / {z}}) \bigr]\,
\nonumber \\
& & \gamma_{\bf q} =  2 \sum_{i=x,y,z}cosq_i \, , \,
\label{eq:5}
\end{eqnarray}

The coordination number $z$ is twice the system dimensionality.

Looking at Eq.~(\ref{eq:3}), it is clear that the unperturbed
Hamiltonian  $H_0$  can be taken as

\begin{eqnarray}
H_0= -\epsilon_p \sum_i n_i + \sum_{\bf q} \omega_{\bf q} b_{\bf q}^{\dag} b_{\bf q}
\label{eq:6} \, , \,
\end{eqnarray}

while the remaining part of the Hamiltonian ($\tilde H-H_0$) in the MLF
basis is considered as the perturbation part.

The energy eigenstates of $H_0$ are given by

\begin{eqnarray}
|\phi_i, \{n_{\bf q}\}\rangle = c_i^{\dag} |0 \rangle_e
|n_{{\bf q}_1}, n_{{\bf q}_2}, n_{{\bf q}_3}, ...\rangle_{ph}
\label{eq:7}
\end{eqnarray}

where, $i$ is the electron site and $n_{{\bf q}_1}, n_{{\bf q}_2},
n_{{\bf q}_3}$ are the phonon occupation numbers in the phonon
momentum states ${\bf q}_1, {\bf q}_2, {\bf q}_3$, respectively.
The lowest energy eigenstate of the unperturbed Hamiltonian has no
phonon excitations, $i.e. \, n_{\bf q}=0$ for all ${\bf q}$. The
ground state has an energy $E_0^0=- \epsilon_p$ and is $N$-fold
degenerate, where $N$ is the number of sites in the system. The
perturbation lifts the degeneracy and to first order in $t$ the
ground state energy of the 3D- polaron with momentum ${\bf k}$ is
given by

\begin{eqnarray}
E_0({\bf k})= - \epsilon_p - t_p \gamma _{\bf k}
\label{eq:8}
\end{eqnarray}

The second order correction to the ground-state energy of the
polaron with momentum ${\bf k}$ is given by

\begin{eqnarray}
E_0^{(2)}({\bf k})& &= \sum_{\bf k'}  \sum_{\{n_{\bf q}\}}
{1 \over {\sum_{\bf q} n_{\bf q} \omega_{\bf q}}} \, \nonumber \\
& & \times
|< \{n_{\bf q}\},{\bf k'}| \tilde {H} - H_0|{\bf k},\{0\}> |^2
\nonumber \\
\label{eq:9}
\end{eqnarray}

where, in principle, intermediate states having all possible phonon numbers contribute to Eq.~(\ref{eq:9}).

By minimizing the zone center ground state energy, the
variational parameters $\lambda_{\bf q}$ are obtained as:

\begin{eqnarray}
\lambda_{\bf q} = {g \over {\omega_{\bf q} + z t_p (1- \gamma_{\bf q} / z)}}
\label{eq:12}
\end{eqnarray}

and, by Eq.~(\ref{eq:12}), the one phonon matrix
element between the ground state $|{\bf k=0},\{n_{\bf
q}=0\}\rangle$ and the first excited state $\langle 1_{\bf q},{\bf k'}|$
vanishes.
Then, the one phonon excitation process yields no
contribution to Eq.~(\ref{eq:9}). By Eq.~(\ref{eq:12}) one gets the $\lambda_{\bf q}$'s for the 1D, 2D
and 3D systems and, via Eq.~(\ref{eq:5}), the narrowing of the polaron band \cite{i104}.

\section*{3. Polaron Mass}

Given the formal background, the polaron mass is calculated both for the Lang-Firsov and for the Modified Lang-Firsov method to the second order in SCPT. Generally the second order order correction {\it i)} makes a relevant contribution to the ground state energy, {\it ii)} does not affect the bandwidth, {\it iii)} introduces the mass dependence on the adiabatic parameter which would be absent in the first order.  Altogether the polaron landscape introduced by the second order SCPT is much more articulated than the simple picture suggested by the first order of SCPT which nonetheless retains its validity towards the antiadiabatic limit. In first order SCPT, ground state energy, bandwidth and effective mass appear as equivalent, interchangeable properties to describe the polaron state while band narrowing and abrupt mass enhancement are {\it coincident} signatures of small polaron formation in parameter space. However, there is no compelling physical reason for such coincidence to occur as the bandwidth is in fact a zone edge quantity whereas the effective mass is a zone center quantity. Thus, by decoupling onset of the band narrowing and self trapping of the polaron mass, the second order of SPCT provides the framework for a richer polaron structure in momentum space.

As emphasized in the Introduction, dispersive phonons have a relevant role in the Holstein model:
a dispersionless spectrum would in fact predict {\it
larger polaron bandwidths in lower dimensionality} and yield a {\it divergent site jump
probability} for the small polaron in time dependent perturbation theory \cite{i55,i86}.
Here, a lattice model is assumed in which first neighbors molecular sites interact through a force constants pair potential.
Then, the optical phonon spectrum is given in 1D, 2D (square lattice) and 3D (simple cubic lattice) respectively, by

\begin{eqnarray}
& &\omega^2_{1D}(q)=\, {{\alpha + \gamma } \over M} + {1 \over M}
\sqrt { \alpha^2 + 2 \alpha  \gamma cosq + \gamma^2}
\nonumber \\
& &\omega^2_{2D}({\bf q})=\, {{\alpha + 2 \gamma } \over M} +
 {1 \over M} \sqrt { \alpha^2 +
2 \alpha  \gamma g({\bf q}) +  \gamma ^2 (2 + h({\bf q}))}
\nonumber \\
& &\omega^2_{3D}({\bf q})=\, {{\alpha + 3 \gamma }
\over M} +
 {1 \over M} \sqrt { \alpha^2 +
2 \alpha  \gamma j({\bf q}) + \gamma ^2 (3 + l({\bf q}))}
\nonumber
\\
& &g({\bf q})=\,cosq_x + cosq_y \nonumber
\\
& &h({\bf q})=\, 2cos(q_x -
q_y)
\nonumber \\
& &j({\bf q})=\,cosq_x + cosq_y + cosq_z
\nonumber
\\
& &l({\bf q})=\, 2cos(q_x - q_y) + 2cos(q_x - q_z) + 2cos(q_y -
q_z)) \nonumber
\\
\label{eq:13}
\end{eqnarray}

where $\alpha$ is the intra-molecular force constant  and $\gamma$ is the
inter-molecular first neighbors force constant. $M$ is the reduced
molecular mass. Thus, the intra- and inter-molecular energies  are
$\omega_0 =\,\sqrt{2\alpha/M}$ and $\omega_1=\,\sqrt{\gamma/M}$ respectively.
Some care should be taken in setting the phonon energies as the second order perturbative term grows faster than the first order by increasing $\omega_1$ \cite{i24}. Hence too large dispersions may cause a breakdown of the SCPT.
In terms of $\omega_0$, the dimensionless parameter $zt/\omega_0$ defines the adiabatic ($zt/\omega_0 > 1$)
and the antiadiabatic ($zt/\omega_0 < 1$) regime. Some other choices are found in the literature with $t/\omega_0 > 1$ or $t/\omega_0 > 1/4$ defining the adiabatic regime. As shown below, such discrepancies may lead to significantly different interpretations of the polaron behavior in parameter space mainly for higher dimensions.

Hereafter I take $\omega_0=100meV$ and tune $t$ to select a set of $zt/\omega_0$ which sample both antiadiabatic and adiabatic regime without reaching the adiabatic limit. The dynamics of the lattice is in fact central to our investigation. A {\it moderate to strong} range of {\it e-ph} couplings is assumed  ($g/\omega_0 \succeq 1$) so that the general conditions for polaron formation are fulfilled throughout the range of adiabatic ratios \cite{i15,i98}.

\begin{figure}
\includegraphics[height=8.0cm,angle=90]{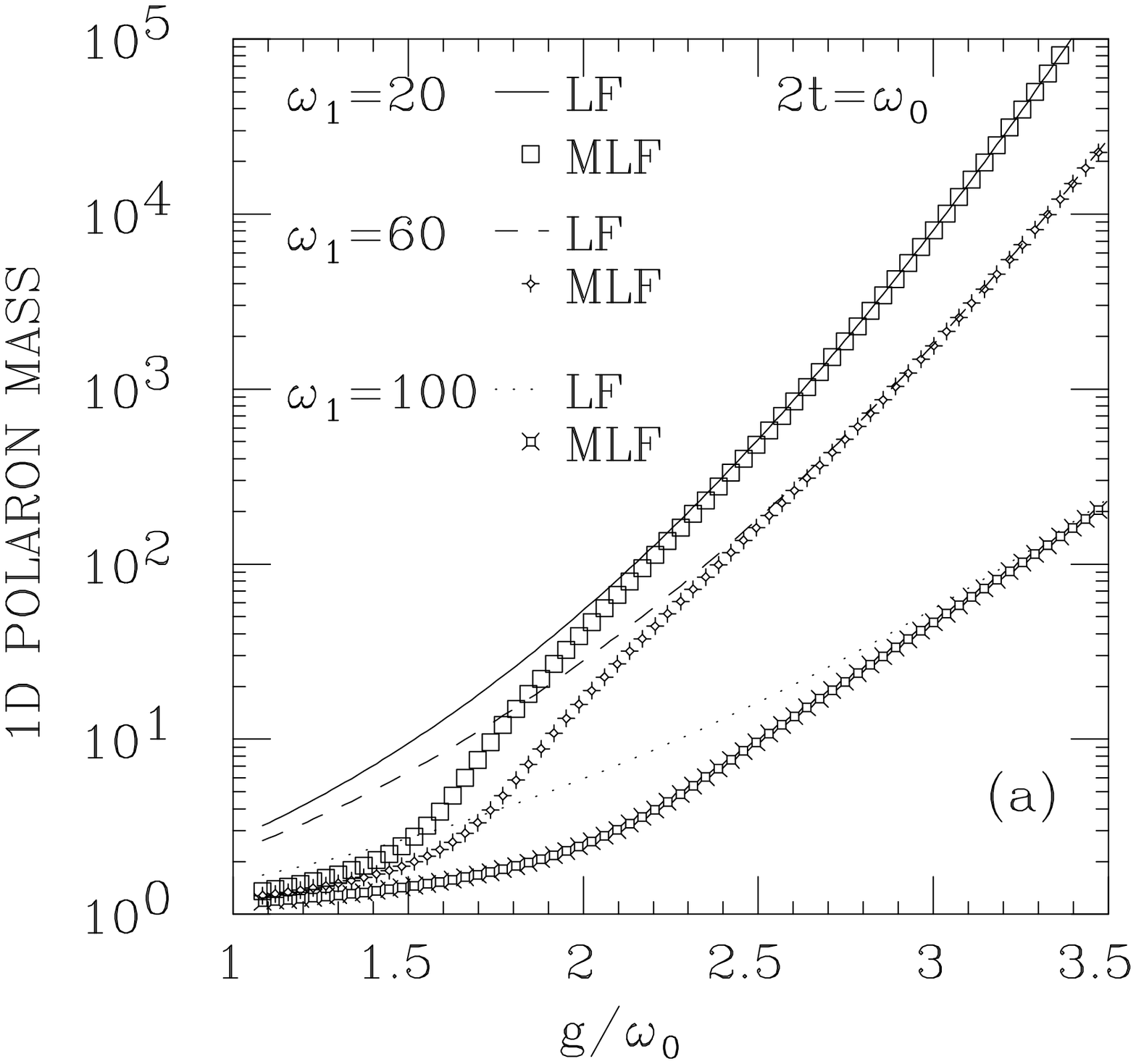}
\includegraphics[height=8.0cm,angle=90]{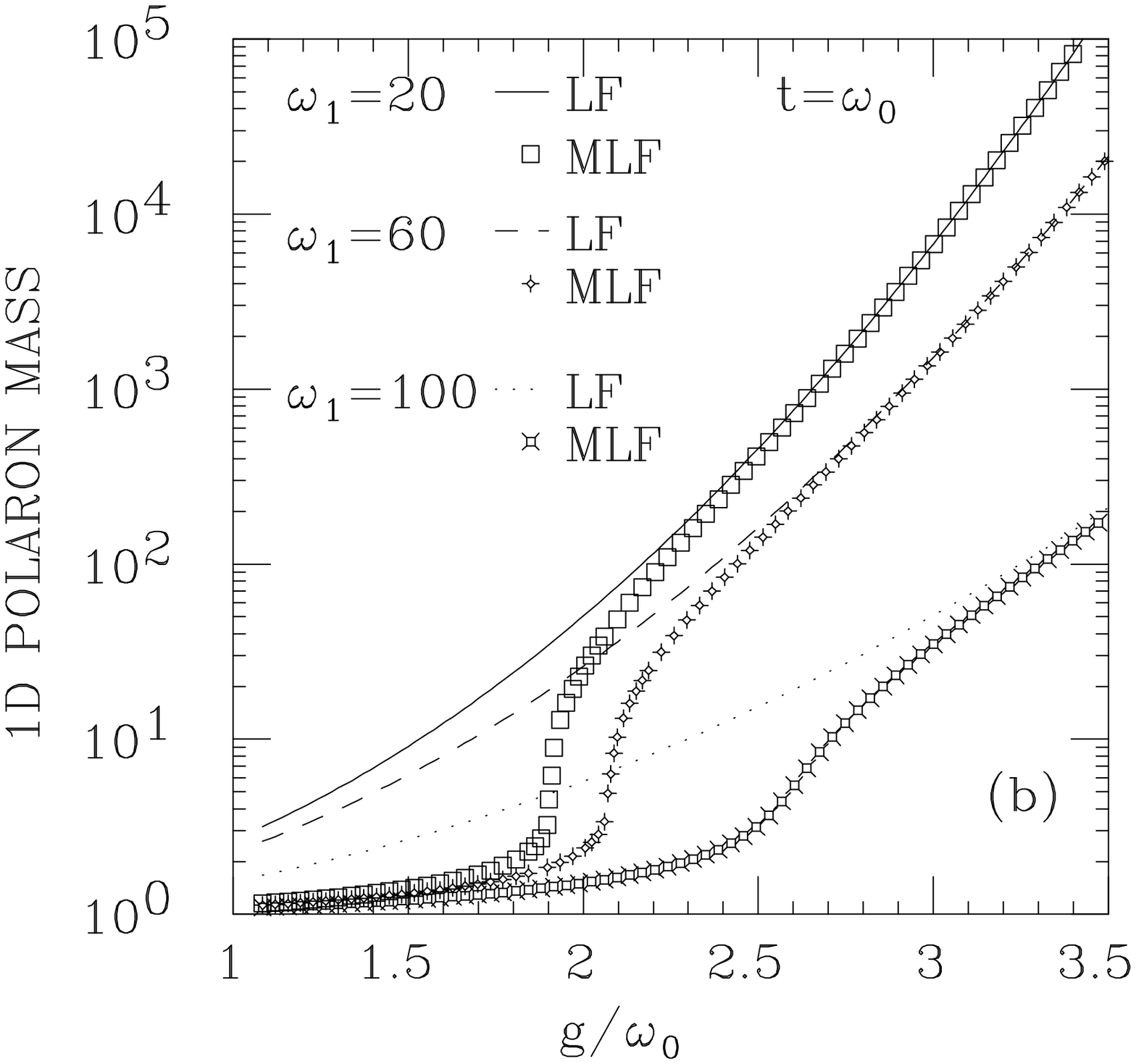}
\includegraphics[height=8.0cm,angle=90]{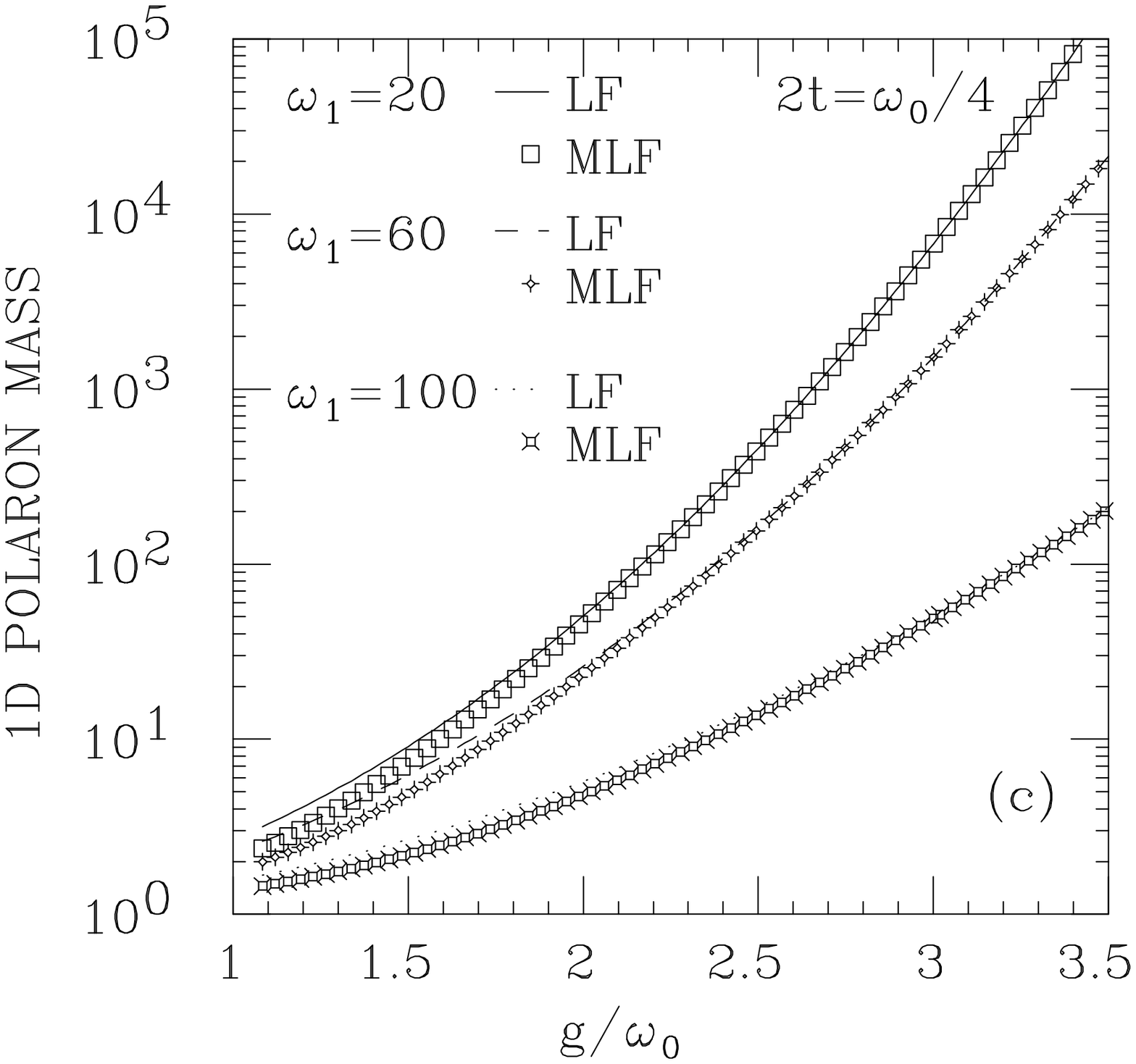}
\caption{\label{fig:1}
Ratio of the one dimensional polaron mass to the bare band mass versus {\it
e-ph} coupling according to the Lang-Firsov and the Modified
Lang-Firsov methods. The adiabatic parameter is set at: (a) the
intermediate value, $2t/\omega_0=\,1$; (b) a fully adiabatic
regime, $2t/\omega_0=\,2$; (c) an antiadiabatic regime,
$2t/\omega_0=\,0.25$. $\omega_0=\,100meV$ and $\omega_1$ (in units
$meV$) are the {\it intramolecular} and {\it intermolecular}
energies of the diatomic molecular chain respectively.}
\end{figure}

Fig.~\ref{fig:1} plots the ratio of the one dimensional polaron mass to the bare band mass, against the {\it e-ph} coupling, calculated both in the Lang-Firsov scheme  and in the Modified Lang-Firsov expression.
The intermediate regime $2t= \omega_0$ is assumed in Fig.~\ref{fig:1}(a).
While at very strong couplings the MLF plots converge towards the LF predictions, a remarkably different behavior between the LF and the MLF mass shows up for moderate $g$.
The LF method overestimates the polaron mass for $g \in [\sim 1 -
2]$ and mostly, it does not
capture the rapid mass increase found instead in the MLF
description. Note that, around the crossover, the MLF polaron mass
is of order ten times the bare band mass choosing
$\omega_1=\,60meV$. Large intermolecular energies enhance the
phonon spectrum thus reducing the effective masses in both
methods. In the MLF method, large $\omega_1$ tend also to smooth
the mass behavior in the crossover region.

Fig.~\ref{fig:1}(b) presents the case of an adiabatic regime: the discrepancies
between LF and MLF plots are relevant for a broad range of
{\it e-ph} couplings.  The latter is more extended for larger intermolecular energies as also seen in Fig.~\ref{fig:1}(a).
Mass renormalization is poor in the MLF curves up
to the crossover which is clearly signalled by a sudden {\it
although continuous} mass enhancement whose abruptness is
significantly smoothed for the largest values of intermolecular
energies.

Fig.~\ref{fig:1}(c) shows a fully antiadiabatic case in which the LF and MLF plots
practically overlap throughout the whole range of
couplings. This confirms that the LF method is appropriate in the antiadiabatic limit which is essentially free from retardation effects. Unlike the previous cases the MLF plots are always smooth versus $g$ indicating that no {\it self-trapping} event occurs in the antiadiabatic regime. Here the polaron does not trap as it is always small for the whole range of couplings.

It should be reminded that the concept of {\it self-trapping} traditionally indicates an
abrupt transition between an infinite size state
at weak {\it e-ph} couplings and a finite (small) size polaron at
strong {\it e-ph} couplings. In one
dimension, according to the traditional adiabatic polaron
theory \cite{i56,i105,i57}, the polaron solution is always the ground state of
the system and no self-trapping occurs. Instead, in higher dimensionality a minimum coupling
strength is required to form finite size polarons, hence
self-trapped polarons can exist at couplings larger than that
minimum.

However, these conclusions have been critically re-examined in the recent polaron literature and the same notion of {\it polaron size} has been questioned due to the complexity of the polaron quasiparticle itself \cite{i81,i59}.
Certainly, as a shrinking of the polaron size yields a weight
increase, the polaron mass  appears as the most
reliable indicator of the self-trapping transition.
The latter, as the results in Fig.~\ref{fig:1} suggest,
is rather a crossover essentially dependent on the degree
of adiabaticity of the system and crucially shaped by the internal
structure of the phonon cloud which I have modelled by tuning the
intermolecular forces. In this view, self
trapping events can be found also in the parameter space of 1D systems. This amounts to say that
also finite size polarons can self-trap if a sudden, although continuous, change in
their effective mass occurs for some values of the {\it e-ph}
couplings in some portions of the intermediate/adiabatic regime. The continuity of such event follows from the fact that the ground state energy is analytic function of $g$ for optical phonon dispersions \cite{i58} hence the possibility of phase transitions is ruled out in the Holstein model. Neither at finite temperature can phase transitions occur as the free energy is smooth for any phonon dispersion \cite{i106,i107}.

\begin{figure}
\includegraphics[height=8.0cm,angle=90]{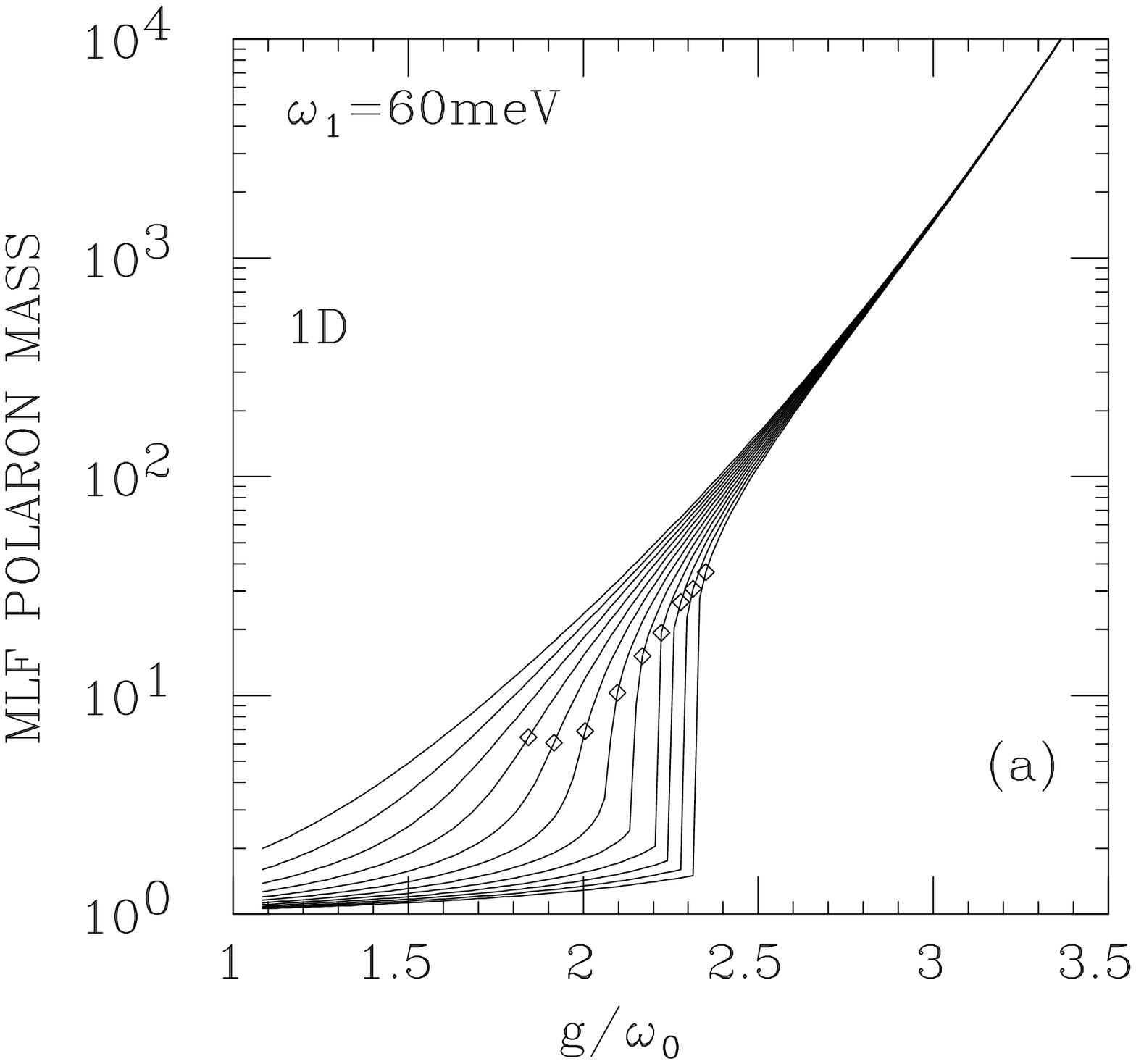}
\includegraphics[height=8.0cm,angle=90]{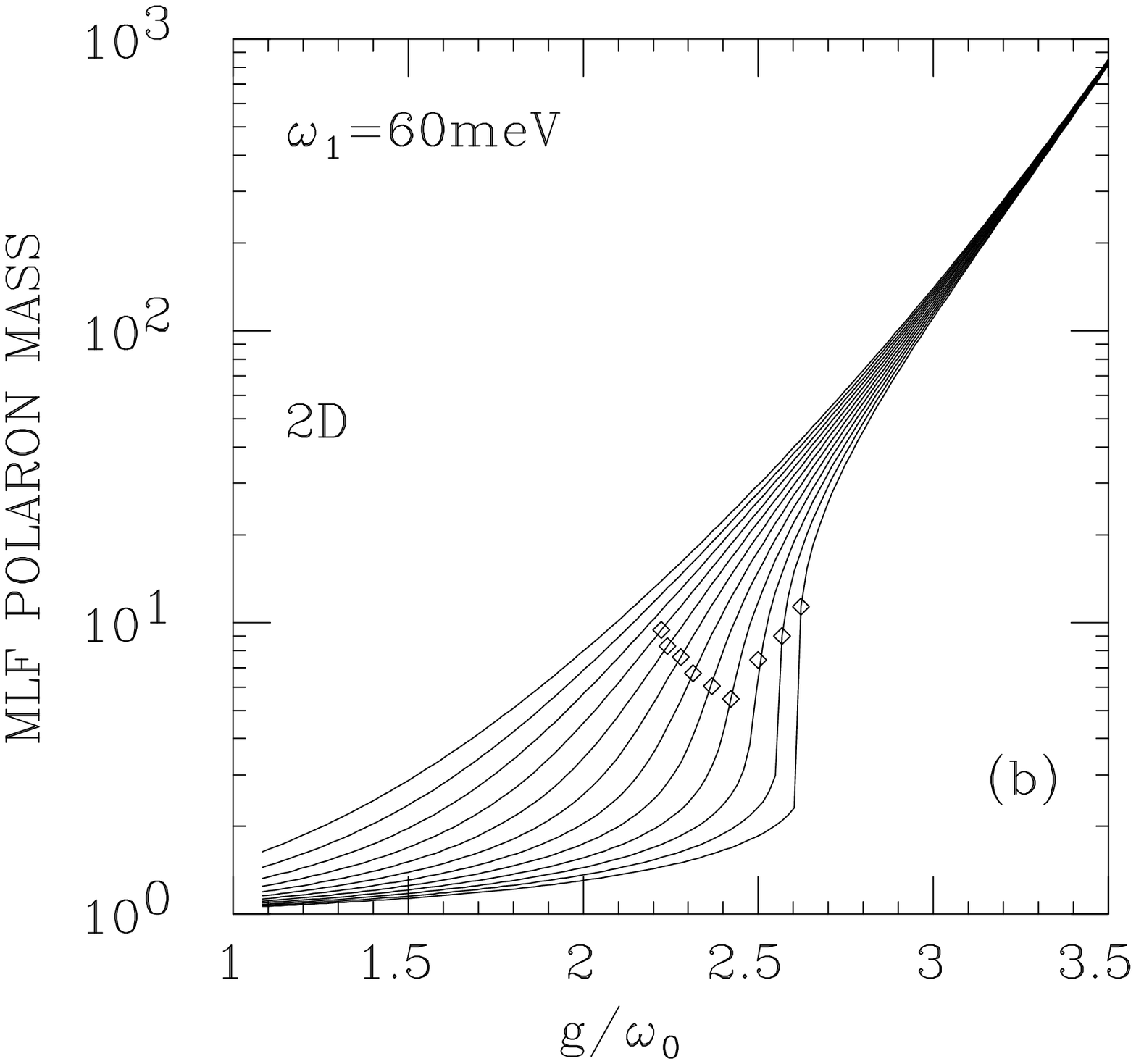}
\includegraphics[height=8.0cm,angle=90]{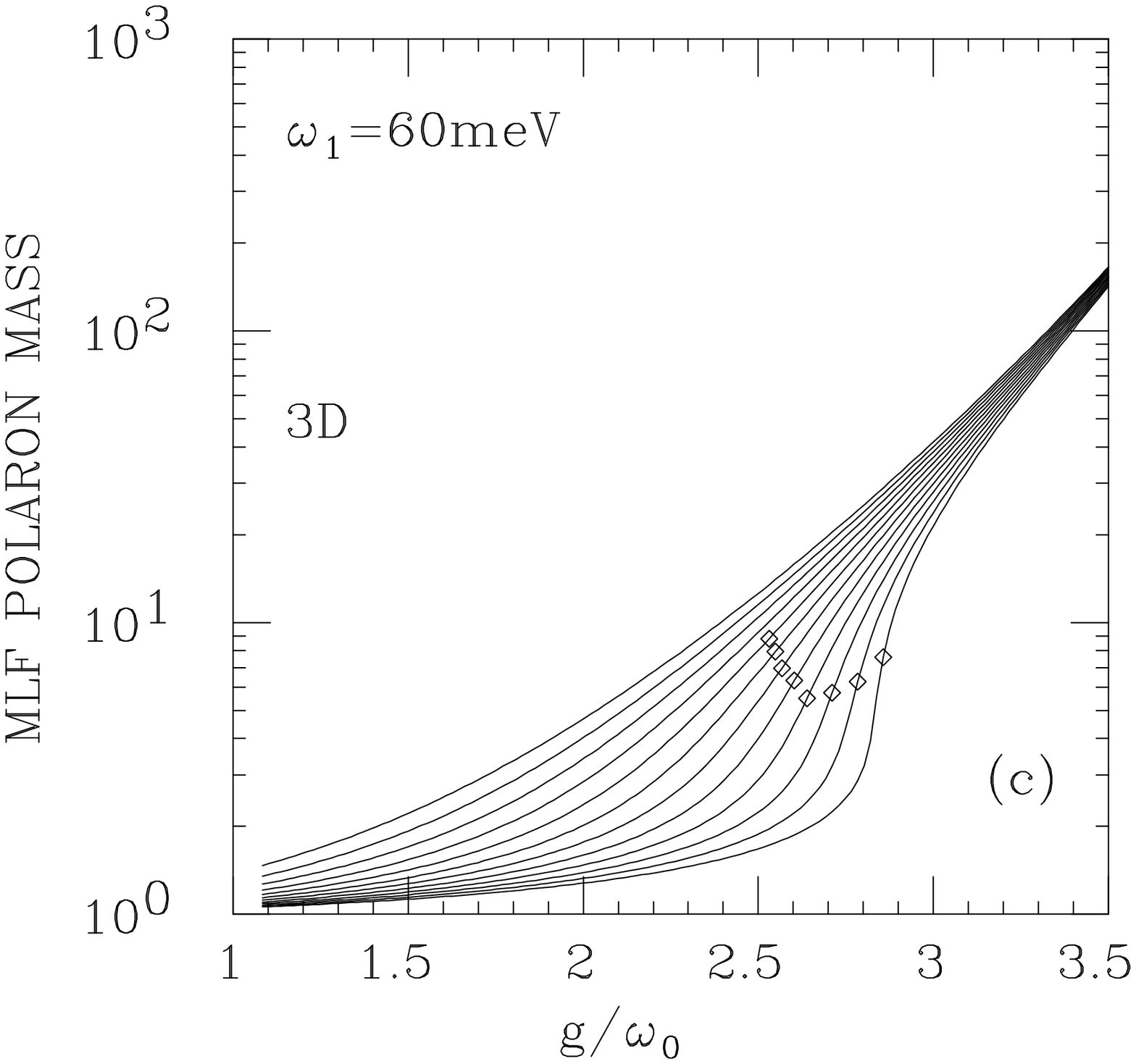}
\caption{\label{fig:2}
Ratio of the Modified Lang-Firsov
polaron mass to the bare band mass versus {\it e-ph} coupling in
(a) 1D, (b) 2D and (c) 3D. A set of twelve $zt/\omega_0$ values
ranging from the antiadiabatic to the adiabatic regime is
considered. From left to right: $zt/\omega_0=\,0.25, 0.5, 0.75,
1.0, 1.25, 1.5, 1.75, 2.0, 2.25, 2.5, 2.75, 3.0$.
$\omega_0=\,100meV$. The diamonds mark the occurence of the
self-trapping event.}
\end{figure}

As fluctuations in the lattice displacements around the electron
site are included in the MLF variational wavefunction, the calculated polaron mass should
not display discontinuities by varying the Hamiltonian parameters  through the crossover \cite{i114}.
Mathematically  the crossover points are selected
through the simultaneous occurence of a maximum in the first
logarithmic derivative and a zero in the second logarithmic
derivative of the MLF polaron mass with respect to the coupling
parameter. Such inflection points, corresponding to the points of
most rapid increase for $m^*$, are marked in Fig.~\ref{fig:2}  on the plots of
the mass ratios computed for a broad range of (anti)adiabatic parameters both in one,
two and three dimensions.

In 1D, see Fig.~\ref{fig:2}(a), the crossover occurs for
$g$ values between $\sim 1.8 - 2.3$ and the corresponding self
trapped polaron masses are of order $\sim 5 - 50$ times the bare band mass
thus suggesting that relatively light small polarons can exist in
1D molecular solids with high phonon spectrum. The onset of the self-trapping line is set
at the intermediate value $2t=\,\omega_0$ and the self-trapped
mass values grow versus $g$ by increasing the degree of
adiabaticity. There is no
self-trapping in the fully antiadiabatic regime as the electron
and the dragged phonon cloud form a compact unit
also at moderate {\it e-ph} couplings. Then, the mass increase
is always smooth in the antiadiabatic regime.

Some significant results are found in 2D as shown in Fig.~\ref{fig:2}(b):
{\it i)} for a given $g$  and adiabaticity ratio, the 2D mass is lighter than
the 1D mass and the 2D LF limit is attained at a value which is
roughly one order of magnitude smaller than in 1D; {\it ii)} the
crossover region is shifted upwards along the $g$ axis with the
self trapping events taking place in the range, $g$ $\sim 2.2 -
2.6$. The masses are of order $\sim 5 - 10$ times
the bare band mass; {\it iii)} the line of
self-trapping events changes considerably with respect to the 1D plot. The marked curve is parabolic in 2D with an extended descending branch starting at
the intermediate value $4t/\omega_0=\,1$; {\it iv)} in the deep
adiabatic regime, the lattice dimensionality smoothens the mass
increase versus $g$.

This effect is even more evident in 3D,
see Fig.~\ref{fig:2}(c), as there are no signs of abrupt mass increase even
for the largest values of the adiabatic parameter. At the
crossover, 3D masses are of order $\sim 5 - 10$ times the bare
band mass with the self trapping points lying in the range, $g$
$\sim 2.5 - 2.9$. At very large couplings the mass ratio becomes independent of $t$  and
converges towards the LF value. In this region (and for the choice
$\omega_1=\,60meV$) the 3D Lang-Firsov mass is one order of
magnitude smaller than the 2D mass. As the coordination number
grows versus dimensionality, large intermolecular forces are more
effective in hardening the 3D phonon spectrum thus leading to
lighter 3D polaron masses than 2D ones.

The self trapping transition appears in Fig.~\ref{fig:2} as smoother in higher dimensions thus contradicting the trend found by previous investigations, markedly by exact diagonalization techniques \cite{i26,i88} and variational calculations \cite{i122}. Two reasons may be invoked to resolve the discrepancy, the first being more physical and the second more technical. {\it 1)} From Eq.~(\ref{eq:13}), it can be easily seen that: $\omega^2_{d}[{\bf q}=\,(0,0,0)] - \omega^2_{d}[{\bf q}=\,(\pi,\pi,\pi)]=\,2 d \omega_1^2$. Then,  for a given $\omega_1$, the phonon band is more dispersive in higher dimensionality $d$. Hence increasing the system dimension corresponds to attribute larger weight to the intermolecular forces which ultimately smoothen the crossover as made evident in Fig.~\ref{fig:1}. {\it 2)} The works in Refs. \cite{i26,i88,i122} take $t/\omega_0$ as adiabatic ratio in any dimension whereas our calculation spans the same range of $zt/\omega_0$ values in any dimension.  Thus, for instance, the  fully adiabatic $zt/\omega_0=\,3$ plots in Fig.~\ref{fig:2} would correspond to an antiadiabatic case in 3D ($t/\omega_0=\,1/2$) and a slightly adiabatic case in 1D ($t/\omega_0=\,3/2$)  following the definition in  Refs. \cite{i26,i88,i122}. Consistently,  the 3D {\it antiadiabatic} polaron mass is smoother than the 1D {\it adiabatic} polaron mass. Viceversa, assuming the same ratio i.e. $t/\omega_0=\,1/2$,  one should compare the rightmost plot in Fig.~\ref{fig:2}(c) with the fourth from left in Fig.~\ref{fig:2}(a): the latter is smoother than the former as the 1D case is now antiadiabatic while the 3D case is fully adiabatic.

Nonetheless, the findings displayed in Fig.~\ref{fig:2} agree with density matrix renormalization group \cite{i21}, variational Hilbert space \cite{i26} methods and weak coupling perturbation theory \cite{i93} in predicting a lighter mass in higher dimensionality.

\section*{4. Electron-Phonon Correlations}

\begin{figure}
\includegraphics[height=9.0cm,angle=90]{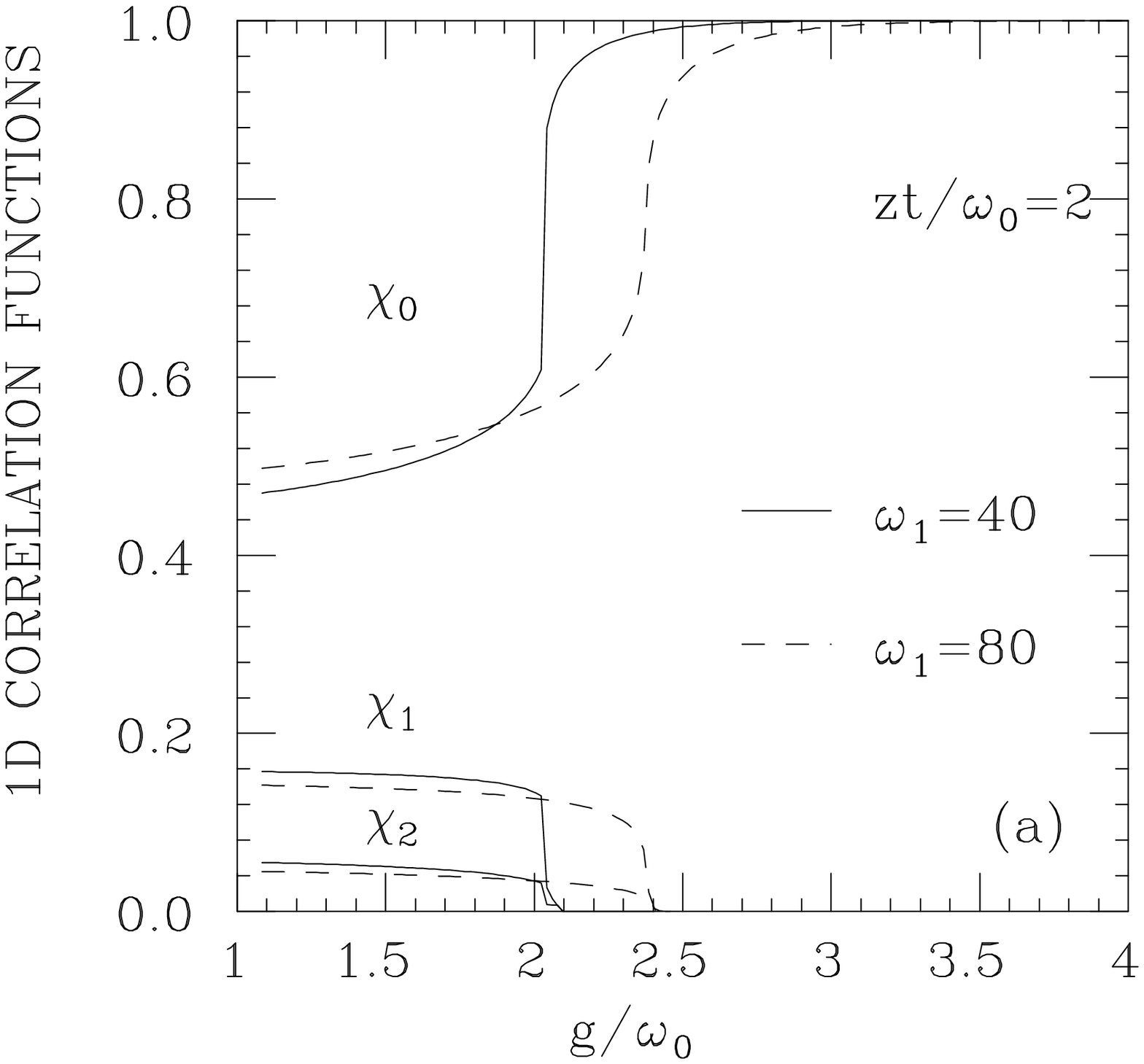}
\includegraphics[height=9.0cm,angle=90]{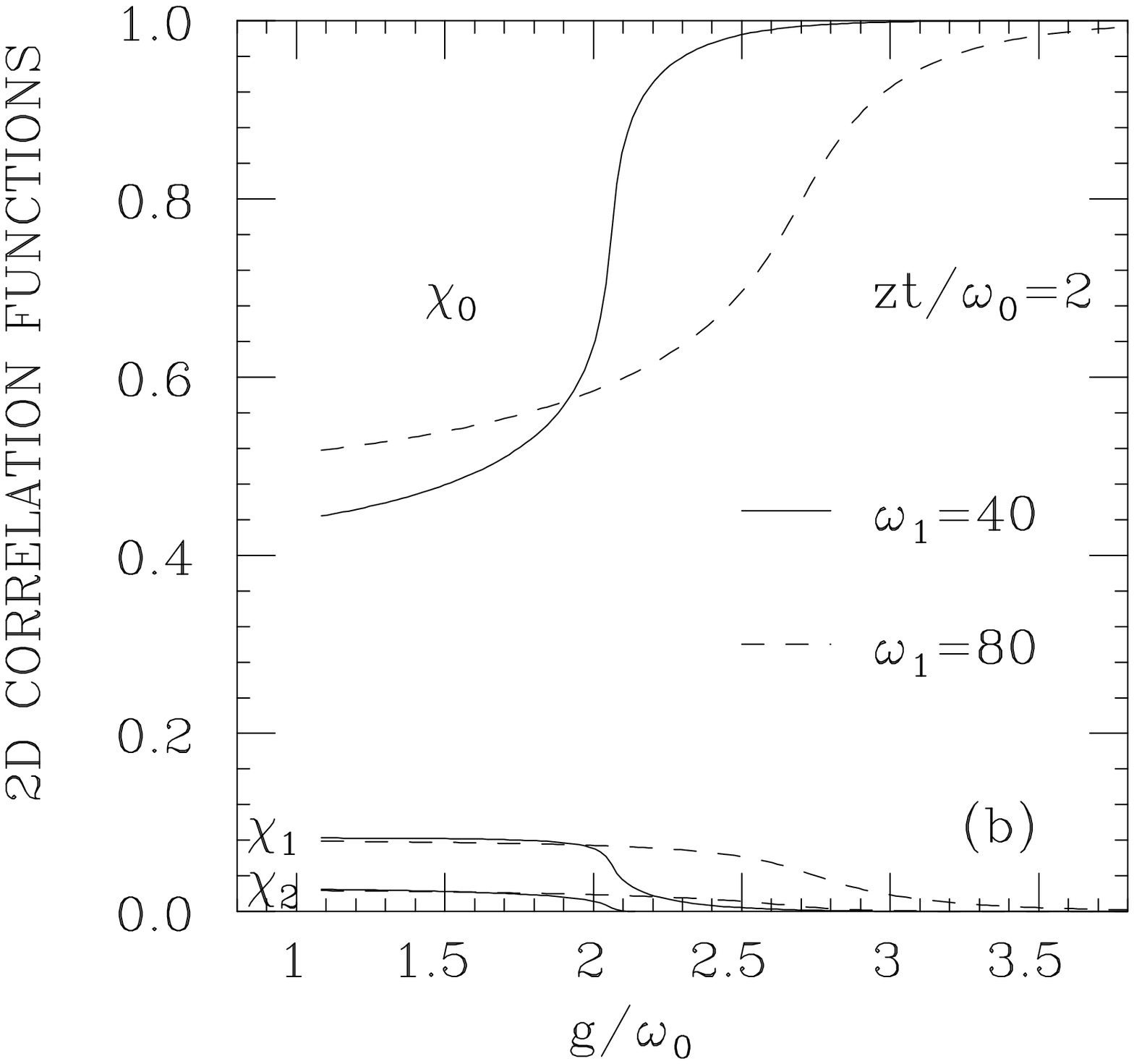}
\caption{\label{fig:3} Electron-phonon correlation functions: on site $\chi_0$, first neighbor site $\chi_1$ and  second neighbor site $\chi_2$ versus {\it e-ph} coupling in
(a) 1D,  (b) 2D.  An adiabatic regime is considered and two values of intermolecular energy are assumed.}
\end{figure}

Within the MLF formalism one may also compute the electron-phonon correlation functions in the polaron ground state. This offers a measure of the polaron size as electron and phonons displacements can be taken at different neighbors sites. The on site  $\chi_0$, the first neighbor site  $\chi_1$ and the second neighbor site  $\chi_2$ correlation functions are plotted in Fig.~\ref{fig:3} against the {\it e-ph} coupling both in 1D and 2D. For $g \sim 1$, there is some residual quasiparticle weight (not appreciable on this scale) associated with $\chi_3$ and $\chi_4$ which however tend to vanish by increasing the coupling. An adiabatic regime is assumed to point out the self-trapping event which occurs when $\chi_0 \sim \,1$ and $\chi_n \sim \,0$ for $n \geq 1$. This happens for $g \sim 2$ in 1D and $g \sim 2.5$ in 2D for the lowest case of $\omega_1$ here considered.
By studying the correlation functions for two values of $\omega_1$ it is seen how the intermolecular forces smooth the crossover and extend the polaron size over a larger range of $g$. The case $\omega_1=\,60meV$, which allows a comparison with Fig.~\ref{fig:2}, is intermediate between the two plots presented in Fig.~\ref{fig:3}. In such case the self trapping appears at  $g \sim 2.3$ in 1D in fair agreement with the corresponding inflection point in the 1D effective mass (see fifth curve from right in Fig.~\ref{fig:2}(a)). In 2D the polaron size shrinks at slightly larger $g$ than in 1D and the on site localization occurs at $g$ somewhat larger than the mass inflection point.

Altogether the diamond marked loci displayed in Fig.~\ref{fig:2} indicate that strong mass renormalization is accompanied by on site (or two sites) polaron localization thus identifying the self trapping line with the formation of small polarons. Before the crossover takes place, for instance at $1 \leq g \leq 2$ in Fig.~\ref{fig:3}(a), the 1D polaron spreads over a few (two to four) lattice sites. This is the also the range of couplings in which the polaron band shows the largest deviations \cite{i90} from the  cosine-like band of standard LF perturbation theory \cite{i53} due to the importance of longer (than nearest neighbors) hopping processes. The notion of {\it intermediate polaron} seems to us as the most appropriate for such mobile and relatively light objects.

\section*{5. Path Integral Method}

The results obtained so far can be put on sound physical bases by applying the space-time path integral method to the dispersive Holstein Hamiltonian. The method permits to incorporate the effect of the electron-phonon correlations in a momentum dependent effective {\it e-ph} coupling.

The phonons operators in Eq.~(\ref{eq:1}) can be generally written in terms of the isotropic
displacement field $u_{\bf n}$ as:

\begin{equation}
b_{\bf i}^{\dag} + b_{\bf i} = {1 \over {N}} \sum_{\bf q}
\sqrt{2M\omega({\bf q})} \sum_{n} \exp(i{\bf q} \cdot ({\bf R}_i
- {\bf R}_n)) u_{n}
\label{eq:14}
\end{equation}

Then, the {\it e-ph} term in Eq.~(\ref{eq:1}) transforms as follows:

\begin{equation}
H^{e-ph} =
      {g \over {N}} \sum_{\bf q} \sqrt{2M\omega({\bf q})}
      \sum_{ i,n} c_{i}^{\dag} c_{i}u_{ n}
      \exp(i{\bf q} \cdot ({\bf R}_i - {\bf R}_n))
\label{eq:15}
\end{equation}

The sum over {$n$} spans all $n^{th}$ neighbors of the
${\bf R}_i$ lattice site in any dimensionality. It is clear that the phonon dispersion introduces {\it
e-ph} real space correlations  which would be absent in a dispersionless model. Note that Eq.~(\ref{eq:15}) contains the same physics as the {\it e-ph} term in Eq.~(\ref{eq:1}).
Fourier transforming the atomic
displacement field and taking the lattice constant $|{\bf
a}|=\,1$, from Eq.~(\ref{eq:15}), I get for a linear chain and a square
lattice respectively:

\begin{eqnarray}
& &H_{d}^{e-ph} =\,
 {g \over {N^{3/2}}}   \sum_{i} c_{i}^{\dag} c_{i}
      \sum_{\bf q, q^{\prime}} \sqrt{2M\omega({\bf q})}
      \exp(i{\bf q^{\prime}} \cdot {\bf R}_i)
\nonumber
\\
& &\times \, u_{\bf q^{\prime}}
S_{d}({\bf q^{\prime} - q})
\nonumber
\\
& &{S}_{1D}({q^{\prime} - q}) \equiv 1 + 2
\sum_{n=1}^{n^*}\cos(n(q^{\prime} - q)) \nonumber
\\
& &S_{2D}({\bf q^{\prime} - q}) \equiv 1 + 2 \sum_{n=1}^{n^*}
\Bigl[\cos (n \Delta q_x) + \cos ( n \Delta q_y)\Bigr]
\nonumber
\\
& &+ \,2  \sum_{m,n=1}^{n^*} \Bigl[ \cos\bigl( m \Delta q_x + n \Delta q_y \bigr)
+ \,\cos\bigl(
m \Delta q_x - n \Delta q_y \bigr) \Bigr]
\nonumber
\\
& &\Delta q_x \equiv \, q^{\prime}_x - q_x ; \, \, \, \Delta q_y \equiv \, q^{\prime}_y - q_y
\nonumber
\\
\label{eq:16}
\end{eqnarray}

While, in principle, the sum over $n$ should cover all the $N$
sites in the lattice, the cutoff $n^*$
permits to monitor the behavior of the coupling term as a function of
the range of the {\it e-ph} correlations. In particular, in 1D the
integer $n$ numbers the neighbors shells up to $n^*$ while in the square lattice,
the $n^*=1$ term includes the second neighbors shell, the sum up
to $n^*=2$ includes the fifth neighbors shells, $n^*=3$ covers the
nineth shell and so on. Switching off the interatomic forces,
$\omega({\bf q})=\,\omega_0$, one would recover from Eq.~(\ref{eq:16}) a local
{\it e-ph} coupling model with $S_d \equiv 1$. As no approximation
has been done at this stage Eq.~(\ref{eq:16}) is general.

\subsection*{A. Semiclassical Holstein Model}

I apply to the Holstein Hamiltonian space-time mapping
techniques \cite{i64,i65,i85} to write the
general path integral for an electron particle in a bath of
dispersive phonons.  The method has been used to treat also {\it e-ph} polymer models \cite{i62,i63} in which the electron hopping causes a {\it shift} in the atomic displacements and, as a consequence, the vertex function
depends both on the electronic and the phononic wave vector. Such {\it shift} is however absent in the Holstein model as Eq.~(\ref{eq:1}) makes clear \cite{i67,i68}.

I introduce ${\bf x}(\tau)$ and ${\bf
y}(\tau')$ as the electron coordinates at the $i$ and $j$ lattice sites respectively, and the electronic Hamiltonian (first term in Eq.~(\ref{eq:1})) transforms into

\begin{eqnarray}
H^e(\tau,\tau')=\, -{t} \bigl( c^{\dag}({\bf x}(\tau)) c({\bf
y}(\tau')) + c^{\dag}({\bf y}(\tau')) c({\bf x}(\tau)) \bigr) \,\nonumber
\\
\label{eq:17}
\end{eqnarray}

$\tau$ and $\tau'$ are continuous variables $\bigl( \in [0,
\beta]\bigr)$ in the Matsubara Green's functions formalism with
$\beta$ being the inverse temperature hence the electron hops are
not constrained to first neighbors sites.  After setting $\tau'=\,0$, ${\bf
y}(0) \equiv 0$, the electron
operators are thermally averaged over the ground state of the Hamiltonian. As a result,
 the average energy per lattice site, $h^e(\tau) \equiv \,<H^e(\tau)> / N$, associated to
electron hopping reads (in $d$ dimensions):

\begin{eqnarray}
& &h^e(\tau) = \, - {t}\Bigl(G[-{\bf
x}(\tau), -\tau ] + G[{\bf x}(\tau), \tau ]\Bigr) \,\nonumber
\\
& &G[{\bf x}(\tau), \tau]=\,{1 \over {\beta }}\int {{d{\bf
k}}\over {\pi}^d} exp[i{\bf k \cdot x}(\tau)]\sum_n
{{exp(-i\hbar\nu_n \tau)} \over {i\hbar\nu_n - \epsilon_{\bf k}}}
\,\nonumber
\\
\label{eq:18}
\end{eqnarray}

$\nu_n$ are the fermionic Matsubara frequencies and $\epsilon_{\bf
k}$ is the electron dispersion relation.

Consider now the {\it e-ph} term.
The spatial {\it e-ph} correlations contained in Eq.~(\ref{eq:16}) are mapped
onto the time axis introducing the $\tau$ dependence in the
displacement field: $u_{\bf q} \to u_{\bf q}(\tau)$. Assuming
periodic atomic particle paths: $u_{\bf q}(\tau + \beta)=\,u_{\bf
q}(\tau)$, the displacement is expanded in $N_F$ Fourier
components:

\begin{equation}
u_{\bf q}(\tau)=\,u_o + \sum_{n=1}^{N_F} 2\Bigl((\Re u_n)_{\bf q}
\cos( \omega_n \tau) - (\Im u_n)_{\bf q} \sin( \omega_n \tau)
\Bigr)\,
\label{eq:19}
\end{equation}

with $\omega_n=\, 2n\pi/\beta$.
I take a semiclassical version of the Holstein Hamiltonian \cite{i76}, assuming that the phonon coordinates in Eq.~(\ref{eq:16}) as classical variables interacting with quantum mechanical fermion operators.  Such approximation may affect the thermodynamics
of the system as the quantum lattice fluctuations in fact play a role mainly for intermediate values of the {\it e-ph} coupling \cite{i69,i75}.

Averaging Eq.~(\ref{eq:16}) on the
electronic ground state, the {\it e-ph} energy per
lattice site is defined as

\begin{eqnarray}
& &{{<H_{d}^{e-ph}>} \over N} = \sum_{\bf q}<H_{d}^{e-ph}>_{\bf q}
\nonumber
\\
& &<H_{d}^{e-ph}>_{\bf q}=\, {g \over {N^{3/2}}}
\sqrt{2M\omega({\bf q})}  \sum_{\bf q^{\prime}} \rho_{\bf
q^{\prime}} u_{\bf q^{\prime}}
S_{d}({\bf q^{\prime} - q})
\nonumber \\
& &\rho_{\bf q^{\prime}}=\,{1 \over N} \sum_{i} <c_{i}^{\dag} c_{i}> \exp(i{\bf q^{\prime}} \cdot {\bf R}_i)
\, \nonumber \\
\label{eq:20}
\end{eqnarray}

Then, on the base of Eq.~(\ref{eq:19})  and Eq.~(\ref{eq:20}),
I identify the perturbing source current \cite{i66} for the Holstein
model with the $\tau$ dependent averaged {\it e-ph} Hamiltonian
term:

\begin{eqnarray}
& &j(\tau)=\, \sum_{\bf q} j_{\bf q}(\tau)\nonumber
\\
& &j_{\bf q}(\tau) \equiv \, <H_{d}^{e-ph}>_{\bf q}. \nonumber
\\
\label{eq:21}
\end{eqnarray}

Note that the Holstein source current does not depend on the
electron path coordinates. The time dependence is incorporated only in the atomic displacements.
This property will allow us to disentangle phonon and electron degrees
of freedom in the path integral and in the total partition
function.

After these premises, one can proceed to write the
general path integral for an Holstein electron in
a bath of dispersive phonons. Assuming a mixed representation, the
electron paths are taken in real space while the phonon paths are
in momentum space. Thus, the electron path integral reads:

\begin{eqnarray}
& &<{\bf x}(\beta)|{\bf x}(0)>=\,\prod_{\bf q}<{\bf x}(\beta)|{\bf
x}(0)>_{\bf q}\, \nonumber
\\
& &<{\bf x}(\beta)|{\bf x}(0)>_{\bf q}=\, \int Du_{\bf q}(\tau)
exp\Biggl[- \int_0^{\beta} d\tau {M \over 2} \bigl( \dot{u_{\bf
q}}^2(\tau)
\nonumber \\ & &+ \omega^2({\bf q}) u_{\bf
q}^2(\tau) \bigr) \Biggr] \,\nonumber
\\  &\times& \int D{\bf x}(\tau) exp\Biggl[- \int_0^{\beta}d\tau
\biggl({m \over 2} \dot{{\bf x}}^2(\tau) + h^e(\tau) - j_{\bf
q}(\tau)\biggr) \Biggr], \nonumber
\\
\label{eq:22}
\end{eqnarray}

where $m$ is the electron mass. The perturbing current in Eq.~(\ref{eq:22}) is integrated over $\tau$  using  Eq.~(\ref{eq:19}) and Eq.~(\ref{eq:21}). The result is:

\begin{eqnarray}
& &\int_0^\beta d\tau j_{\bf q}(\tau)=\,{\beta u_o} {g}_d({\bf q})
\,\nonumber \\ & &{g}_d({\bf q})=\, {g \over {N^2}}
\sqrt{2M\omega({\bf q})} \sum_{\bf q^{\prime}} \rho_{\bf
q^{\prime}} S_{d}({\bf q^{\prime} - q}) \nonumber \\
\label{eq:23}
\end{eqnarray}

where ${g}_d({\bf q})$ is thus a time averaged {\it e-ph} potential.

The total partition function can be derived from Eq.~(\ref{eq:22}) by imposing
the closure condition both on the phonons (Eq.~(\ref{eq:19})) and on the
electron paths, ${\bf x}(\beta)=\,{\bf x}(0)$. Using Eq.~(\ref{eq:23}), I
obtain:

\begin{eqnarray}
& &Z_T=\,\prod_{\bf q} \oint Du_{\bf q} \exp\Biggl[{\beta
u_o} {g}_d({\bf q}) - \int_0^{\beta} d\tau {M \over 2} \bigl(
\dot{u_{\bf q}}^2(\tau)\,  \nonumber
\\
& &+\, \omega^2_{\bf q} u_{\bf q}^2(\tau)
\bigr) \Biggr] \,\nonumber
\\
& &\times \oint D{\bf x} \exp\Biggl[-
\int_0^{\beta}d\tau \biggl({m \over 2} \dot{{\bf x}}^2(\tau) +
h^e(\tau) \biggr) \Biggr]  \nonumber
\\
\label{eq:24}
\end{eqnarray}

$\oint Du_{\bf q}$ and $\oint D{\bf x}$ are
the measures of integration which normalize
the kinetic terms in the phonon field and electron actions respectively  \cite{i66}.

The phonon degrees of freedom in Eq.~(\ref{eq:24}) can be integrated out analytically yielding:

\begin{eqnarray}
& &Z_{T}=\,\prod_q P({\bf q}) \times \oint D{\bf x}(\tau) \, \nonumber \\
& &\times \,
exp\Biggl[- \int_0^{\beta}d\tau \biggl({m \over 2} \dot{\bf
x}^2(\tau) + h^e(\tau) \biggr) \Biggr] \, \nonumber \\
& &P({\bf q})= \,{1 \over {\beta \omega({\bf q})}}\exp
\Biggl[{{\bigl({g}_d({\bf q}) \lambda_M \bigr)^2} \over {2\pi
\omega({\bf q})^2}} \Biggr] \, \nonumber \\
& &\times \,
\prod_{n=1}^{N_F} {{(2n\pi)^2} \over
{(2n\pi)^2 + (\beta \omega({\bf q}) )^2}} \,
\nonumber \\
\label{eq:25}
\end{eqnarray}

being $\lambda_M=\,\sqrt{\pi \hbar^2 \beta/M}$.

Eq.~(\ref{eq:25}) is the final analytical result from which the thermodynamics of the model can be computed \cite{i68}.
The exponential function in $P({\bf q})$ embodies the effect of the
non local correlations due to the dispersive nature of the phonon
spectrum. Phonon and electron contributions to the partition
function are decoupled although the effective potential $g_d({\bf
q})$ carries a dependence on the electron density profile in
momentum space through the function $\rho_{\bf q}$.

\begin{figure}
\includegraphics[height=9.0cm,angle=90]{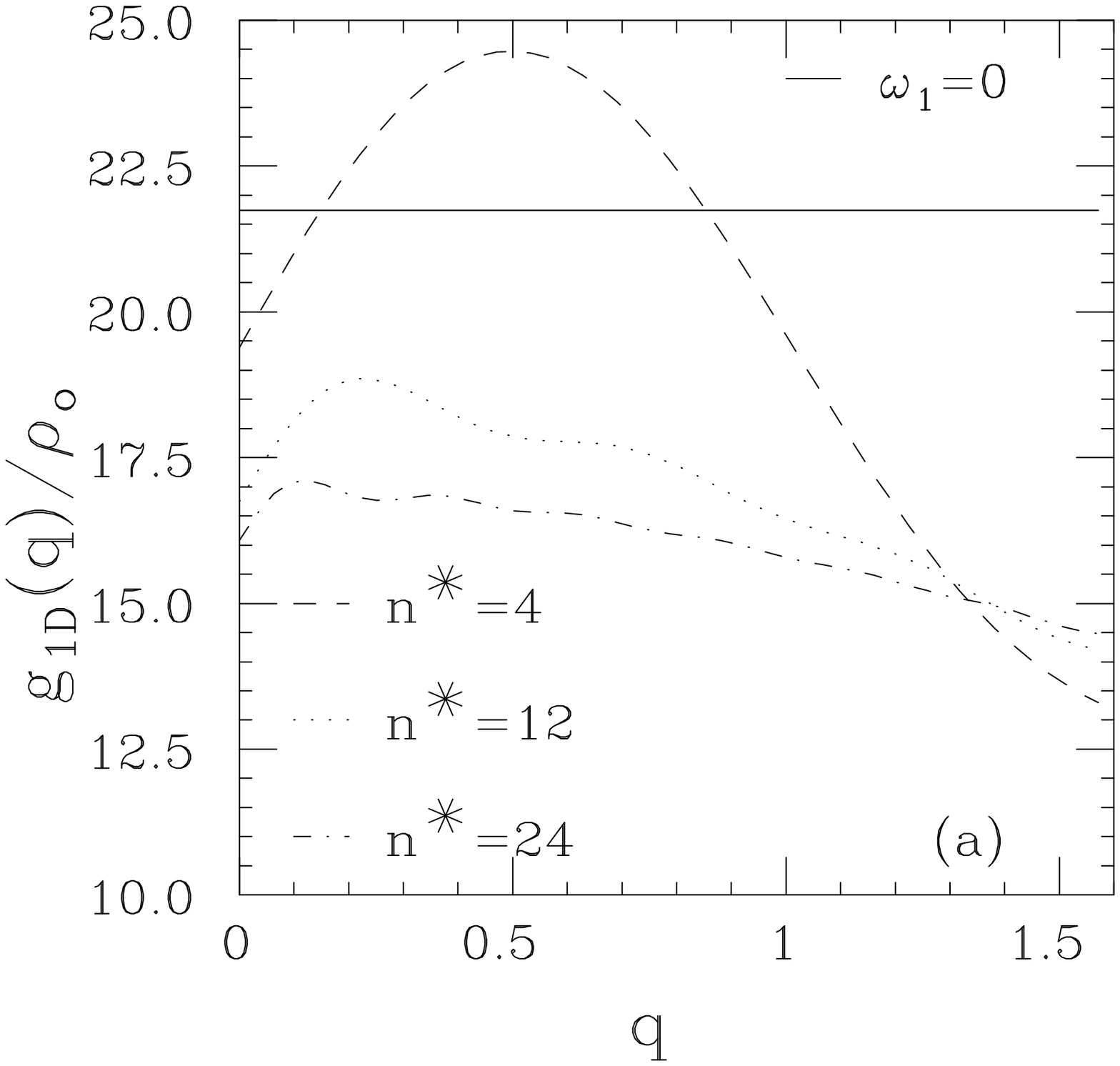}
\includegraphics[height=9.0cm,angle=90]{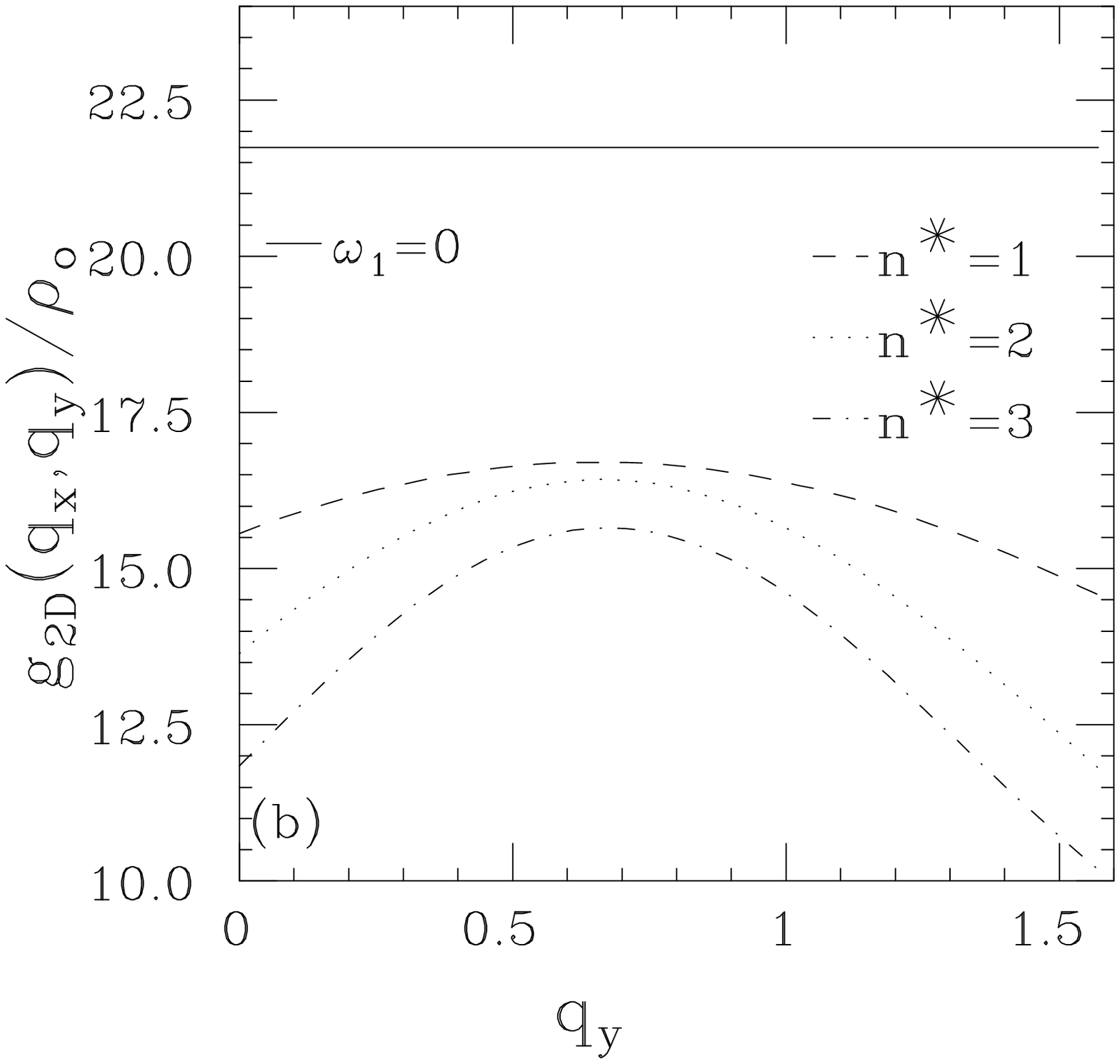}
\caption{\label{fig:4}
(a) Time averaged {\it e-ph} coupling (in units $meV
\AA^{-1}$ versus wave vector for a linear chain. $n^*$ represents
the cutoff on the {\it e-ph} correlations. $\rho_0$ is the
electron density. $\omega_0=20meV$ and $\omega_1=10meV$. $g=3$.
The dispersionless {\it e-ph} coupling is obtained for
$\omega_1=0$. (b) Time averaged {\it e-ph} coupling in two
dimensions versus the $y$ component of the momentum. The cases
$n^*=1$, $n^*=2$ and $n^*=3$ imply that the correlation range
includes the second, the fifth and the nineth neighbors shell,
respectively. The input parameters are as in (a).}
\end{figure}

\subsection*{B. Electron-Phonon Coupling}

The time (temperature) averaged
{\it e-ph} potential in Eq.~(\ref{eq:23}) is computed in the case of a linear chain and
of a square lattice. As an example, I take the electron density profile as $\rho_{q}=\,\rho_o
\cos(q)$ in 1D and $\rho_{{\bf q}}=\,\rho_o
\cos(q_x)\cos(q_y)$ for the 2D system. Since the momentum
integration runs over $q_i \in [0,\pi/2]$, $\rho_o$ represents in
both cases the total electron density. More structured density
functions containing longer range oscillations are not expected to
change the trend of the results hereafter presented.
Low energy phonon spectra parameters are here assumed, $\omega_0=20meV$ and
$\omega_1=10meV$. Setting $g=\,3$, I take a strong Holstein
coupling  although the general trend of the results holds for
any value of $g$.

In Fig.~\ref{fig:3}(a), ${{ g}_{1D}({q})/{{\rho_o}}}$ is
plotted for three choices of the cutoff $n^*$  to
emphasize how the potential depends on the {\it e-ph}
interaction range. The constant value of the potential obtained
for $\omega_1=\,0$ is also reported on.  For short
correlations ($n^*=\,4$) there is a range of wave vectors in which
the effective coupling becomes even larger than in the dispersionless
case: this is due to the fact that the approximation related to a short cutoff is not sufficiently accurate. In fact, extending the range of the {\it e-ph} correlations,
the effective potential is progressively and substantially reduced for all momenta
with respect to the dispersionless Holstein model.
Numerical convergence for ${{ g}_{1D}({q})}$ is found at the value
$n^*=\,24$ which corresponds to 48 lattice sites along the chain.
Then, the 1D coupling renormalization is
q-dependent and generally larger for $q \sim \pi/2$ where the phonon dispersion generates stronger {\it e-ph} correlations. Exact diagonalization techniques \cite{i90} had found that multiphonon states with longer range hopping processes substantially soften the polaron band narrowing with respect to the predictions of standard SCPT with the largest deviations occuring at the polaron momentum $\sim \pi/2$. In fact strong, nonlocal {\it  e-ph} correlations should favour hopping on further than neighbors sites occupied by phonons belonging to the lattice deformation.

The projections of the two dimensional {\it e-ph} potential along
the $y$ component of the wave vector is shown in Fig.~\ref{fig:3}(b)
for three values of the cutoff. For
$n^*=\,1$ the correlation range is extended to the second neighbor
shell thus including 8 lattice sites. For $n^*=\,2$ and $n^*=\,3$
the normalization is over 24 and 48 lattice sites respectively. Then, the
three values of $n^*$ in 2D span as many lattice sites as the
three values of $n^*$ in 1D respectively. I have made this choice to
normalize consistently the potential for the linear chain and the
square lattice. There is a strong renormalization for the 2D
effective potential with respect to the dispersionless case for
any value of the wave vector. Remarkably, the reduction is already evident for $n^*=\,1$.
By extending the correlation range
this tendency becomes more pronounced for $q_y$ close to the
center and to the edge of the reduced Brillouin zone. An analogous
behavior is found by projecting the 2D potential along the $q_x$
axis. The 2D potential stabilizes by including the $9th$ neighbor
shell ($n^*=\,3$) in the correlation range.

Then, an increased range for the {\it e-ph} correlations leads to
a very strong reduction of the effective
coupling, an effect which is qualitatively analogous to that one would get
by hardening the phonon spectrum. This is the physical reason underlying the fact that the Holstein polaron mass can be light.

\section*{6. Conclusions}

According to the traditional notion of small polaron, the strong electron coupling to the lattice deformation implies a polaron collapse of the electron bandwidth ($zt$) and remarkable mass renormalization. Thus, bipolaronic theories have been thought of being inconsistent with  superconductivity at high $T_c$ as the latter is inversely proportional to the bipolaron effective mass.
Motivated by these issues, I have investigated the conditions for the existence of light polarons in the Holstein model and examined the concept of self-trapping versus dimensionality for a broad range of (anti)adiabatic regimes. The self trapping events correspond to the points of most rapide increase for the polaron effective mass versus the {\it e-ph} coupling. A modified version of the Lang-Firsov transformation accounts for the spreading of the polaron size due to retardation effects which are relevant mainly for moderate {\it e-ph} couplings. Assuming large optical phonon frequencies ($\omega_0$), it is found that polaron masses display an abrupt although continuous increase driven by the {\it e-ph} coupling in any dimension, included the 1D case. Such self trapping events occur in the adiabatic and intermediate ($zt/\omega_0 \sim 1$) regimes provided that the {\it e-ph} coupling is very strong ($g/\omega_0 \geq 2$).
The second order of strong coupling perturbation theory permits to decouple two fundamental properties of the polaron landscape, namely band narrowing and effective mass enhancement. I have considerd a range of {\it e-ph} couplings ($ g/\omega_0 \geq 1$) in which the conditions for polaron formation are fulfilled. For a given value of the adiabatic ratio, the onset of the band narrowing anticipates, along the $g$ axis, the crossover to the heavy polaron state which pins the quasiparticle essentially on one lattice site. In such buffer zone, between band narrowing onset and self trapping, the charge carriers can be appropriately defined as {\it intermediate polarons}.

In all dimensions, there is room for non trapped intermediate polarons in the region of the moderate couplings. Such polarons spreads over a few lattice sites and their real space extension grows with dimensionality. Thus, analytical methods can suitably describe some polaron properties also in that interesting intermediate window of {\it e-ph} couplings (i.e. $1 \leq g/\omega_0 \leq 2$ for the 1D system) in which adiabatic polarons are not large but not yet on site localized.
In the square lattice, at the onset of the self trapping state, the {\it polaron over bare electron} mass ratios are $\sim 5 - 10$ and even smaller for lower couplings.

The physical origins for the possibility of light polarons have been analysed treating the dispersive Holstein Hamiltonian model by the imaginary time path integral method. The perturbing source current is the time averaged {\it e-ph} Hamiltonian term. The partition function of the model has been derived integrating out the phonon degrees of freedom in semiclassical approximation.
Focusing on the role of the intermolecular forces, I have quantitatively determined the effects of the dispersive spectrum on the time-averaged {\it e-ph} coupling which incorporates the nonlocality effects.
It is found that, in the square lattice, the {\it e-ph} coupling is strongly renormalized downwards and this phenomenon is already pronounced by the (minimal) inclusion of the second neighbors intermolecular shell. This may explain why 2D Holstein polarons can be light while the real space range of the {\it e-ph} correlations is relatively short.

\section*{Acknowledgements}

Joyous and fruitful collaboration with Prof. A.N. Das is acknowledged.

\end{document}